\documentstyle[twocolumn,prl,aps,epsfig]{revtex}
\begin{document}                                                              
\newcommand{\be}{\begin{eqnarray}}
\newcommand{\ee}{\end{eqnarray}}
\title{Hadron production in nuclear collisions at RHIC and high density QCD}
\author{Dmitri Kharzeev$^a$ and Marzia Nardi$^b$}

\bigskip
 
\address{
a) Department of Physics,\\ Brookhaven National Laboratory,\\
Upton, New York 11973-5000, USA\\
b) Dipartimento di Fisica Teorica dell'Universit{\`a} di Torino and INFN, Sezione di Torino\\ 
via P.Giuria 1, 10125 Torino,
     Italy}
\date{\today}
\maketitle
\begin{abstract} 
We analyze the first results on charged particle multiplicity at RHIC 
in the conventional eikonal approach and in the framework of high density QCD. 
We extract the fraction $F$ of the hadron multiplicity 
originating from ``hard'' (i.e. proportional to the number of binary collisions) processes;
we find a strong growth of this fraction with energy: $F(\sqrt{s}=56\ {\rm GeV}) \simeq 22 \%$, while  
 $F(\sqrt{s}=130\ {\rm GeV}) \simeq 37 \%$.     
This indicates a rapid increase in the density of the produced particles.    
We outline the predictions of high density QCD for the centrality, energy, and atomic number dependence 
 of hadron production. Surprisingly, the predictions of the conventional eikonal approach and of high 
density QCD for centrality dependence of hadron multiplicity at $\sqrt{s}=130\ {\rm GeV}$ appear very similar.

\end{abstract}
\begin{narrowtext}

\section{Introduction}

Recently, the first results of the  charged multiplicity measurement at RHIC 
were published by the PHOBOS Collaboration \cite{Phobos}. These data marked 
the beginning of collider era in the experiments with relativistic heavy ions, 
and already excited a controversy about the 
mechanism of multi-particle production at this new high--energy frontier 
\cite{WG,Kari,KJ,theory,Kahana}. 
One of the focusing points of the discussion has been the r{\^o}le of saturation 
effects \cite{Kari,WG}.

The purpose of the present paper is twofold. First, we analyze the experimental 
results in the framework of 
the eikonal Glauber approach which has been shown to describe well the data 
on particle production in heavy ion collisions at SPS energies (
a detailed description of the formalism and further references can be found in Ref. \cite{KLNS}).  
We then extract the fraction of the hadron multiplicity originating from the processes scaling 
with the number of binary collisions (``hard processes'') and make prediction for the centrality 
dependence. Second, we discuss whether or not 
the present data indicate the onset of saturation behavior, 
expected in high--density QCD \cite{hdQCD,MV,AHM}. 
We briefly review the physical picture of parton saturation and make  
predictions for the centrality, energy, and atomic number dependence of particle 
production in nuclear collisions; we then compare these to the predictions based on the 
conventional eikonal approach.   

It has been conjectured long time ago \cite{bm} that at sufficiently high energy 
the particle production in nucleus--nucleus collisions will be dominated by 
hard processes. However, the gross features of particle production at CERN SPS energies 
were found to be approximately consistent (see, e.g., \cite{exp}) with the scaling in the number 
of participants $N_{part}$, accommodated by the 
``wounded nucleon model'' \cite{Bialas}\footnote{(Multi)strange 
particle yields \cite{WA97} show a remarkable 
disagreement with this trend, but here we will address only the total charged multiplicity.}.  
There are deviations from 
this scaling; however they are not very large -- the WA98 Collaboration, for example,
finds \cite{WA98} the charged particle pseudo-rapidity density at mid-rapidity scaling as $\sim N_{part}^{1.08}$.
The advent of RHIC has pushed the energy of heavy ions to the new frontier, and the 
first results on charged multiplicity \cite{Phobos} at two different energies 
($\sqrt{s} = 56\ {\rm GeV}$ and 
$\sqrt{s} = 130\ {\rm GeV}$) could already enable us to evaluate the relative importance 
of hard processes in heavy ion collisions at collider energies. 

The PHOBOS Collaboration presented a  
comparison of their results with the multiplicity measured in $\bar{p}p$ and $pp$ 
collisions, 
which shows that the particle pseudo-rapidity density per participant increases 
by approximately $70\%$ near 
pseudo-rapidity $\eta=0$ in $Au-Au$ collisions at $\sqrt{s} = 130\ {\rm GeV}$. This is indicative 
of a significant contribution from hard processes\footnote{While it is true that hard processes 
scale with $N_{coll}$, there may also be a component 
of the ``soft'' interaction with the same scaling, see e.g. \cite{dpm,soft}; the multiplicity analysis 
alone cannot distinguish between these two options, but they would lead to different predictions 
for the transverse momentum distributions.}. We will now quantify this conclusion using 
the eikonal approach.

\section{The eikonal approach to hadron production}
 
Let us assume that the fraction $x$ of the multiplicity $n_{pp}$ 
measured in $pp$ collisions per unit of (pseudo)rapidity 
is due to ``hard'' processes, with the remaining fraction $(1-x)$ being from ``soft'' 
processes. The multiplicity in nuclear collisions will then also have two components: 
``soft'', which we assume is proportional to the 
number of participants $N_{part}$, and ``hard'', 
which is proportional to the number of binary collisions $N_{coll}$:
\be
{dn \over d \eta} = (1-x)\ n_{pp} \frac{\left< N_{part} \right>}{2} + x\  n_{pp} 
\left< N_{coll} \right>. 
\label{xmult}
\ee
The same functional dependence has been used by Gyulassy and Wang in their analysis \cite{WG} 
based on the HIJING model.  

PHOBOS Collaboration quotes the following fit  
to the data on the pseudo-rapidity density of charged multiplicity 
in non-single diffractive $\bar{p}p$ interactions \cite{Abe}: $n_{pp} = 2.5 - 0.25 \ln(s) + 0.023 
\ln^2(s)$. When used at $\sqrt{s}= 130$ GeV, this formula gives $n_{pp}(\sqrt{s} = 130\ {\rm GeV}) 
\simeq 
2.25$; at $\sqrt{s} = 56\ {\rm GeV}$ one gets $n_{pp}(\sqrt{s} = 56\ {\rm GeV}) \simeq  
1.98$. These are the values that will be used as an input to (\ref{xmult}) in this paper.  
 
The shape of the multiplicity distribution at a given (pseudo)rapidity $\eta$ can 
now be readily obtained in the usual way (see, e.g., \cite{KLNS}):
\be
\frac {d \sigma} {d n} = \int d^2b \ {\cal P}(n;b)\ (1 - P_0(b)),  
\ee
where $P_0(b)$ is the probability of no interaction among the nuclei at a given 
impact parameter $b$: 
\be
P_0(b) = (1 - \sigma_{NN} T_{AB}(b))^{AB}, 
\ee
where  
$\sigma_{NN}$ is the inelastic nucleon--nucleon cross section, and $T_{AB}(b)$ is the 
nuclear overlap function for the collision of nuclei with atomic numbers A and B; 
we have used the three--parameter Woods--Saxon nuclear density distributions \cite{tables}.
For $\sqrt{s} = 56\ {\rm GeV}$ we have used $\sigma_{NN}=37 \pm 1 \ {\rm mb}$, while 
for $\sqrt{s} = 130\ {\rm GeV}$  $\sigma_{NN}=41 \pm 1 \ {\rm mb}$ was chosen 
basing on the interpolation of existing $pp$ and $\bar{p}p$ data 
\cite{PDG}.   
For $\sqrt{s} = 200\ {\rm GeV}$, we use 
$\sigma_{NN}=42 \pm 1 \ {\rm mb}$. 
The correlation function ${\cal P}(n;b)$ is given by
\be
{\cal P}(n;b) = \frac{1}{\sqrt{2\pi a \bar{n}(b)}}\ \exp\left( - \frac{(n - \bar{n}(b))^2}{2 a 
\bar{n}(b)}\right),
\ee
here $\bar{n}(b)$ is the mean multiplicity at a given impact parameter $b$, evaluated 
analogously to Eq.(\ref{xmult}); the formulae for the number of participants and the number of binary 
collisions can be found in \cite{KLNS}.  The parameter 
$a$ describes the strength of fluctuations; we have chosen the value $a = 0.6$ which 
fits the data well. 
 
In Fig.1, we compare the 
resulting distributions for two values of $x=0$ and $x=0.09$ to the PHOBOS 
experimental distribution, measured in the interval $3 < |\eta| < 4.5$. 
(The reason for presenting the result corresponding to this particular 
value of $x$ will become clear later; 
we have performed calculations for a wide range of $x$.). 
One can see that both $x=0$ and $x=0.09$ distributions describe the data quite well; 
the curve with $x=0$ (pure wounded nucleon model) 
fits the data somewhat better. This suggests that the multiplicity 
{\sl{ in the $3 < |\eta| < 4.5$ 
pseudo-rapidity interval}} is due to the ``soft'' processes. This is in accord both with 
the perturbative QCD picture, where the minijet production is peaked at mid--rapidity, 
and with the soft production approaches \cite{dpm,soft}.  
On the other hand, 
since the shape of the distribution is described well in our approach and is 
not very sensitive to the relative 
contributions of hard and soft components (we have checked that this is true up to 
$x \simeq 0.12$), we can use this distribution as 
a good handle on centrality.   

The total {\it Au--Au} cross section computed in our approach at $\sqrt{s}=130$ A GeV is 
$\sigma_{tot} = 7.04 \pm 0.05$ barn.

\begin{figure}[b]
\epsfig{file=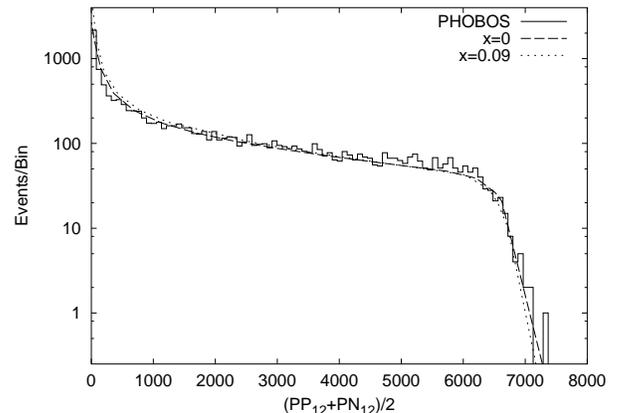, width=3.2in}
\caption{Charged multiplicity distribution at $\sqrt{s}=130$ A GeV; solid line (histogram) -- 
PHOBOS result {\protect\cite{Phobos}}; dashed line -- distribution corresponding to 
participant scaling ($x=0$); 
dotted line -- distribution corresponding to the $37 \%$ admixture of ``hard'' component 
in the multiplicity ($x=0.09$); see 
text for details.}
\label{mult}
\end{figure}

Let us now establish the correspondence between centrality and the mean number of participants.  
The mean number of participants in a nucleus--nucleus collision with 
multiplicity $n$ is defined as 
\be
N_{part}(n) = {\frac{\int d^2 b\  {\cal P}(n;b) (1 - P_0(b)) N_{part}(b)}
{\int d^2 b\  {\cal P}(n;b) (1 - P_0(b))}},
\ee   
where  $N_{part}(b)$ is the mean number of participants 
in the Glauber approach. 
Analogous formula can be used to evaluate the mean number of
collisions 
$N_{coll}(n)$
(for explicit expressions, we refer to \cite{KLNS}).

If we now apply the $6\% $ centrality cut (the cut applied by the PHOBOS Collaboration 
to extract the multiplicity in most ``central'' collisions in the pseudo-rapidity 
interval $| \eta |<1$) to the distribution of Fig. 1 computed with $x=0$, 
we can extract the mean number of participants and the mean number of collisions in the 
events with this centrality: 
\be
\left< N_{part} \right>_{n>n_0} = {\frac{ \int\limits_{n_0} d n \int d^2 b\  {\cal P}(n;b) 
(1 - P_0(b)) N_{part}(b)}
{\int\limits_{n_0}^{\mbox{ }} d n \int d^2 b\  {\cal P}(n;b) (1 - P_0(b))}}, 
\ee
where $n>n_0$ corresponds to the $6\%$ centrality cut in nuclear collisions.

The numbers deduced in this way are the following:
\be 
 \left< N_{part} \right> &=& 339 \pm 2; \nonumber \\
 \left< N_{coll} \right> &=& 1026 \pm 6; \quad\quad \ \sqrt{s} = 130\ {\rm GeV}. \label{npart1}
\ee    
and
\be 
 \left< N_{part} \right> &=& 334 \pm 2; \nonumber \\
 \left< N_{coll} \right> &=& 921 \pm 6; \quad\quad \ \sqrt{s} = 56\ {\rm GeV}. \label{npart2}
\ee 
The numbers of participants (\ref{npart1}, \ref{npart2}) 
are consistent with the ones quoted by the PHOBOS Collaboration: 
$\left< N_{part} \right> = 343 \pm 4(stat)^{+7}_{-14}(syst)$ 
for $\sqrt{s} = 130\ {\rm GeV}$ and  
$\left< N_{part} \right> = 330 \pm 4(stat)^{+10}_{-15}(syst)$ for  $\sqrt{s} = 56\ {\rm GeV}$.

For the sake of completeness, we give in Table 1 the results of the calculations 
of mean number of participants and collisions for different fractions of the 
cross section and different energies. One can extract from this Table also 
the mean numbers corresponding to different cuts on centrality; for completeness, we list them 
in Table 2 together with the corresponding average impact parameters and the mean densities 
of participants in the transverse plane for $\sqrt{s} = 130\ {\rm GeV}$.  

\bigskip

Table 1. The mean number of participants and binary collisions corresponding 
to different centrality cuts in $Au Au$ collisions at different energies, 
as computed in Glauber approach.

\bigskip
\bigskip
\begin{tabular}{c@-c@{\%~}|cc|cc|cc}
\multicolumn{2}{c|}{centr.} 
 &\multicolumn{2}{c|}{$\sqrt{s}=56\  {\mathrm GeV}$} 
 &  \multicolumn{2}{c|}{$\sqrt{s}=130\ {\mathrm GeV}$} 
 &  \multicolumn{2}{c}{$\sqrt{s}=200\ {\mathrm GeV}$} \\
\multicolumn{2}{c|}{cut}
   & $\left< N_{part} \right>$ & $\left< N_{coll} \right>$
   & $\left< N_{part} \right>$ & $\left< N_{coll} \right>$
   & $\left< N_{part} \right>$ & $\left< N_{coll} \right>$ \\ \hline
0 & 5 &  342  & 949  & 344 & 1053 & 344  & 1074 \\
0 & 6 &  336  & 927  & 339 & 1028 & 339  & 1049 \\
0 & 10&  313  & 847  & 316 &  937 & 317  & 958  \\
0 & 20&  266  & 684  & 268 &  755 & 270  & 776  \\
0 & 30&  228  & 563  & 231 &  622 & 230  & 634  \\
0 & 40&  196  & 469  & 198 &  516 & 199  & 528  \\
0 & 50&  169  & 393  & 172 &  434 & 172  & 444  \\
0 &100&  92  & 206  & 93  & 226  &  93  & 231  \\
\hline
10& 20&  218  & 522  & 221 & 575  & 222  & 590  \\
20& 30&  150  & 317  & 153 & 348  & 153  & 356  \\
30& 40&  101  & 184  & 102 & 201  & 102  & 204  \\
40& 50&   63  &  98  &  64 & 106  &  64  & 108  \\
50& 60&   37  &  47  &  38 & 52   &  38  &  52  \\
60& 70&   19  &  21  &  20 & 22   &  20  &  22  \\
70& 80&  8.7  &  7.8 & 9.4 & 8.8  & 9.5  & 8.8  \\
80&100&  2.5  &  1.9 & 2.7 & 2.2  & 2.8  & 2.2  \\

\end{tabular}
\bigskip

We are now in a position to evaluate the ``soft'' and ``hard'' contributions to 
the observed multiplicity by using Eq. (\ref{xmult}) and the experimental 
values of $dn/d\eta = 555 \pm 12(stat) \pm 35(syst)$ at $\sqrt{s} = 130\ {\rm GeV}$ 
and $dn/d\eta = 408 \pm 12(stat) \pm 30(syst)$ at $\sqrt{s} = 56\ {\rm GeV}$. 
Using $n_{pp} = 2.25$ as follows from the fit to the data \cite{Abe}, we get $x = 0.09 \pm 0.03$ 
at $\sqrt{s} = 130\ {\rm GeV}$; the use of the same fit at $\sqrt{s} = 56\ {\rm GeV}$ 
yields $n_{pp} = 1.98$ and leads to the value $x = 0.05 \pm 0.03$. 
These numbers are qualitatively consistent with the predictions 
based on the mini-jet picture (see \cite{Wang} for review), which indicate that the mini-jet 
contribution is increasing as a function of energy between 
$\sqrt{s} = 56\ {\rm GeV}$ and $\sqrt{s} = 130\ {\rm GeV}$.

Using Eq.(\ref{xmult}), we find that at  $\sqrt{s} = 56\ {\rm GeV}$ the fraction  
$F \simeq 22\%$ of the produced particles result from hard processes, 
while at $\sqrt{s} = 130\ {\rm GeV}$ this fraction increases to  $F \simeq 37\%$; 
we define $F$ as
\be
F = {x \; n_{pp} \; N_{coll} \over dn/d\eta}. 
\ee
\bigskip

Table 2. The impact parameter dependence of the mean number of participants and binary collisions 
in $Au Au$ system at $\sqrt{s} = 130$ GeV; we also list the average 
densities of participants in the transverse plane and the corresponding values of saturation scale 
$Q_s$ in the high density QCD approach computed according to Eq.(\ref{satexact}). 

\bigskip

\begin{tabular}{c|c|c|c|c}
$b$ & $N_{part}$ & $N_{coll}$ & $\rho_{part}$ & $Q_s^2$ \\
\footnotesize{(fm)}&            &            &
   \footnotesize{(fm$^{-2}$)} & \footnotesize{(GeV$^2$)} \\
\hline
 0. & 378.4 & 1202.7 & 3.06 & 2.05 \\
 1. & 372.4 & 1173.6 & 3.05 & 2.04 \\
 2. & 354.7 & 1092.9 & 3.01 & 2.02 \\
 3. & 327.0 &  975.7 & 2.95 & 1.98 \\
 4. & 292.2 &  837.0 & 2.86 & 1.92 \\
 5. & 253.2 &  689.6 & 2.75 & 1.84 \\
 6. & 212.3 &  543.5 & 2.60 & 1.74 \\
 7. & 171.5 &  406.7 & 2.41 & 1.61 \\
 8. & 132.4 &  285.6 & 2.17 & 1.46 \\
 9. &  96.5 &  184.8 & 1.89 & 1.26 \\
10. &  65.1 &  107.4 & 1.54 & 1.04 \\
11. &  39.5 &   54.1 & 1.15 & 0.77 \\
12. &  20.5 &   22.7 & 0.72 & --   \\
13. &   8.7 &    7.8 & 0.35 & --   \\
14. &   2.9 &    2.2 & 0.12 & --   \\
15. &   0.8 &    0.5 & 0.03 & --   \\
\end{tabular}

\bigskip

While we have shown that the shape of the minimum bias distribution is not very 
sensitive to the relative magnitude of ``soft'' and ``hard'' contributions, 
the correlation between the forward energy and the multiplicity produced at 
mid-rapidity is \cite{KLNS}. To demonstrate this, we plot the forward energy 
$E_F = (A - N_{part}/2) \sqrt{s}/2$ versus the multiplicity Eq. (\ref{xmult}) in 
Fig. {\ref{zcal}}. However if one normalizes the purely ``soft'' contribution to the 
measured central multiplicity, the difference in the correlations becomes 
less pronounced -- see Fig. {\ref{zcal}}. 

One can see that the correlation found in the case of the 
wounded nucleon model ($x=0$) is significantly different from the correlation 
with $x=0.09$, which corresponds to the fraction $F \simeq 37 \%$ of multiplicity produced 
in hard processes. Since at RHIC the zero degree calorimeter detects only 
the energy carried by neutral particles, the translation of the forward energy 
to the energy measured by the ZDC requires a dedicated study.

\begin{figure}[b]
\epsfig{file=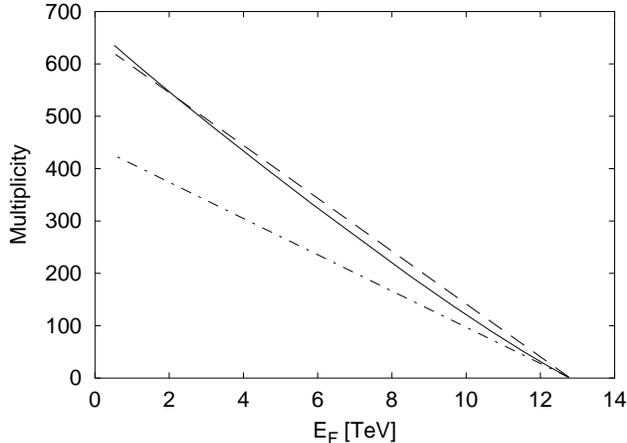, height=2.4in}
\caption{Correlation between the charged multiplicity near $\eta = 0$ and the forward 
energy; dash--dotted line -- the correlation corresponding to the participant scaling; 
solid line -- the correlation containing $37\%$ admixture of ``hard'' component in the 
multiplicity; dashed line is the participant scaling correlation normalized to the measured 
central multiplicity.}
\label{zcal}
\end{figure}

\begin{figure}[b]
\epsfig{file=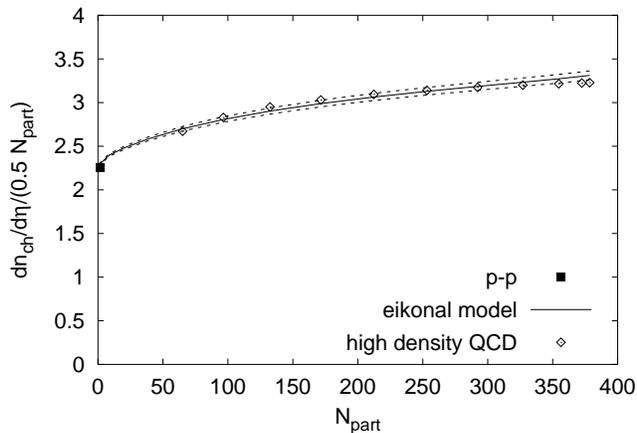, height=2.4in}
\vskip0.1cm
\caption{Centrality dependence of the charged multiplicity per participant pair near $\eta = 0$ at 
$\sqrt{s} = 130$ A GeV; the curves represent the prediction based on the conventional eikonal 
approach, while the diamonds correspond to the high density QCD prediction (see text). The square 
indicates the $pp$ multiplicity.}
\label{centr}
\end{figure} 

The formula (\ref{xmult}) and the calculated numbers of participants and collisions given 
in Tables 1 and 2 allow us to compute also the centrality dependence of the multiplicity. 
With the parameters described above, formula (\ref{xmult}) leads to 
\be
{2 \over N_{part}} {dn \over d\eta} \simeq {1 \over N_{part}} 
\left(2.04\ N_{part} + 0.40\ N_{coll}\right). \label{numcon}
\ee
The result 
is shown in Fig. (\ref{centr}). It is qualitatively, but not quantitatively, consistent 
with predictions of the HIJING model presented in \cite{WG}.

A qualitatively different behavior has been predicted in the framework of 
the saturation approach of Ref. \cite{Kari}, in which the multiplicity per 
participant {\it decreases} as a function of centrality. The approach\cite{Kari} 
assumes the saturation of the {\it produced} partons, whereas the original 
saturation ideas \cite{hdQCD,MV} concern the behavior of partons in the {\it initial} wave 
function of the nucleus. While the two approaches bear some similarities, 
they also have important differences. These differences, as we will see, result in qualitatively different 
predictions for the centrality dependence of particle multiplicity. 
It is therefore worthwhile to examine the predictions 
of the (initial state) saturation approach. 

\section{High density QCD and hadron production in nuclear collisions}

Let us begin by sketching 
the basic ideas of saturation \cite{hdQCD,MV,AHM} 
in their most rudimentary form. Consider the wave function 
of the nucleus boosted to a large momentum. The nucleus is Lorentz--contracted, and partons 
``live'' on a thin sheet in the transverse plane. 
Each parton occupies the transverse area $\pi/Q^2$ determined, 
by uncertainty principle, by its transverse momentum $Q$, and can be probed with the cross section 
$\sigma \sim \alpha_s(Q^2)\ \pi/Q^2$. On the other hand, the entire transverse area of the nucleus 
is $S_A \sim \pi R_A^2$. Therefore, if the number of partons exceeds 
\be  
N_A \sim {S_A \over \sigma} \sim {1 \over \alpha_s(Q^2)} \ Q^2 R_A^2, \label{sat}
\ee
they will begin to overlap in the transverse plane and start interacting with each other, 
which prevents further growth of parton densities\footnote
{Closely related to this picture are the ideas of string percolation \cite{perc,perc1} in the 
transverse plane.}. This happens 
when the transverse momenta of the partons are on the order of 
\be 
Q_s^2 \sim \alpha_s(Q_s^2)\ {N_A \over R_A^2} \sim A^{1/3}, \label{satscale}
\ee
which is called the ``saturation scale''. In the saturation regime, as is apparent from (\ref{sat}) 
and (\ref{satscale}), the multiplicity of the produced partons should be proportional to 
\be
N_s \sim {1 \over \alpha_s(Q_s^2)} \ Q_s^2 R_A^2 \sim N_A \sim A. \label{satmult}
\ee
 In the weak coupling regime, the density of partons becomes 
very large, which justifies the semi--classical McLerran--Venugopalan 
approach \cite{MV}. In the first approximation, the multiplicity in this high density 
regime scales with the number of participants.
There is, however, an important logarithmic correction to this from 
the evolution of parton structure functions with the saturation scale $Q_s^2$, which we discuss below. 

While our ``derivation'' has been very simplistic, 
the formulae (\ref{satscale},\ref{satmult}) can be reproduced by more sophisticated 
methods \cite{hdQCD,MV,AHM}, which also allow to reconstruct the coefficient of 
proportionality in (\ref{satscale}):
\be
Q_s^2 = {8 \pi^2 N_c \over N_c^2 - 1}\ \alpha_s(Q_s^2)  \ xG(x,Q_s^2) \ {\rho_{part}\over 2}, \label{satexact}
\ee 
where $N_c=3$ is the number of colors, $xG(x,Q_s^2)$ is the gluon structure function of the nucleon, and 
$\rho_{part}$ is the density of participants in the transverse plane.  
We divide $\rho_{part}$ by $2$ to get the density of those nucleons in a single nucleus which will participate 
in the collision at a given impact parameter. 

Let us estimate the saturation scale from (\ref{satexact}) 
for a central $Au-Au$ collision at $\sqrt{s} = 130$ GeV. Eq.(\ref{satexact}) is an equation that can be 
solved by iterations; a self--consistent solution can be found at $Q_s^2 \simeq 2\ {\rm GeV}^2$ if we use  
$xG(x,Q_s^2) \simeq 2$ \cite{MRST} at $x\simeq 2Q_s/\sqrt{s}\simeq 0.02$, with $\alpha_s(Q_s^2) \simeq 0.6$.

 As before, we will normalize 
the prediction to the experimental data referring to the $6 \%$ of most central collisions. We use an explicit 
expression \cite{AHM} for the number of produced partons   
\be
{d^2 N \over d^2b d \eta}  = c\ {N_c^2 - 1 \over 4 \pi^2 N_c}\ {1 \over \alpha_s}\  Q_s^2,
\ee
where $c$ is the ``parton liberation'' coefficient accounting for the transformation of virtual partons in the 
initial state to the on--shell partons in the final state. 
Integration over the transverse coordinate and the use of (\ref{satexact}) yield simply 
\be
{dN \over d \eta} = c\ N_{part} \ xG(x, Q_s^2). \label{mults}
\ee
If we {\it assume} that ${dN / d \eta} \simeq 3/2\ {dn_{exp} / d\eta}$, 
take $xG(x,Q_s^2) \simeq 2$ at $Q_s^2 \simeq 2\ {\rm GeV}^2$ and $x\simeq 2Q_s/\sqrt{s}\simeq 0.02$ \cite{MRST}, 
and use $N_{part} \simeq 339$ from Table 1 for the $6 \%$ centrality cut, the experimental number \cite{Phobos}  
 $dn/d\eta = 555 \pm 12(stat) \pm 35(syst)$ translates into the following value of the 
``parton liberation'' coefficient:
\be
c = 1.23 \pm 0.20. \label{val}
\ee
This number appears to be close to unity, as expected by Mueller \cite{AHM}, 
which implies a very direct correspondence between the number of 
the partons in the initial and final states. Moreover, this may imply that the number of particles 
is conserved through the parton--to--hadron transformation -- a miraculous fact first noted in the context 
of the ``local parton--hadron duality'' hypothesis \cite{Dok}.

The value (\ref{val}) can be compared to the recent lattice calculation \cite{KV} by Krasnitz and Venugopalan, 
which yields $c = 1.29 \pm 0.09$. Very recently, an analytical calculation for $c$ has been presented by 
Kovchegov \cite{Yura}, with the result $c = 2 \ln 2 \simeq 1.39$.

To compute the  
centrality dependence, we still need to know the evolution of the gluon structure function 
with the density of partons, which is proportional to the mean density of participants 
in the transverse plane. We will assume that this evolution is governed by the DGLAP equation \cite{DGL,AP}, 
and take 
\be
x G(x, Q_s^2) \sim \ln{\left(Q_s^2 \over \Lambda_{QCD}^2\right)}. \label{log}
\ee 
The dependence (\ref{log}) emerges when the radiation of gluons is treated classically, and so is 
consistent with Eq. (\ref{satexact}). 
Eqs. (\ref{satexact}) and (\ref{mults}) can now be used to evaluate the centrality dependence; with the 
parameters described above we get
\be
{2 \over N_{part}}\ {dn \over d \eta} \simeq 0.82\ \ln{\left(Q_s^2 \over \Lambda_{QCD}^2\right)}, 
\label{satrel}
\ee
where we take $\Lambda_{QCD}\simeq 200$ MeV and use the values of $Q_s$ listed in Table 2.

We present the results of our calculations in the saturation scenario in Fig. \ref{centr}.
The similarity of the predictions based on the conventional eikonal approach and on the high density QCD 
is striking in spite of the {\it a priori} totally different functional dependences (\ref{numcon}) and 
(\ref{satrel}).

\section{Summary and discussion}

To summarize, our analysis of the first RHIC results \cite{Phobos} implies  
that the r{\^o}le of hard processes in particle production rapidly 
increases at collider energies, bringing us to the regime of high parton densities. 
The experimental study of centrality and atomic number dependence of particle 
production at different energies is needed to extract the crucial information about the 
behavior of QCD at these conditions. 

We have made predictions for the centrality 
dependence of the particle multiplicity both in conventional eikonal approach and in the 
framework of high density QCD. The latter prediction is different from the ones available 
in the literature \cite{Kari,WG} and shows that the multiplicity per participant increases 
as a function of centrality, reflecting the evolution of parton densities with the 
increasing saturation scale. Surprisingly, both conventional and high density QCD approaches 
lead to almost identical dependence on centrality. This makes it difficult to 
distinguish between the two approaches using only the data on centrality dependence of 
multiplicity at one fixed energy, and one has to rely on the analysis of the transverse momentum 
distributions. We leave this for the future. On the other hand, the numerical similarity of the  
predictions of eikonal and high density QCD approaches means that the apparent success of the conventional 
approach in describing the previously available data might mask a different physics. 

The approach based on the saturation allows us to extract the initial energy density of partons 
achieved in $ Au Au$ collisions at RHIC. Indeed, in this approach the formation time of 
partons is $\tau_0 \simeq 1/Q_s$, and the transverse momenta of partons are about $k_t \simeq Q_s$. 
We thus can use the Bjorken formula \cite{Bj} and the set of parameters deduced above to estimate
\be
\epsilon \simeq {<k_t> \over \tau_0} \ {d^2 N \over d^2b d \eta} \simeq Q_s^2 \ {d^2 N \over d^2b d \eta} \simeq 
18\ {\rm {GeV/fm^3}}.
\ee
This value is well above the energy density needed to induce the  
QCD phase transition according to the lattice calculations.  

High density QCD also provides very definite predictions for the energy and atomic number 
dependence of hadron multiplicity -- according to (\ref{mults}), the multiplicity simply 
scales with the number of participants and the gluon structure function of the nucleon, which 
one has to evolve to the saturation scale. The knowledge of the extracted value (\ref{val}) 
furnishes the prediction. 
 
We are fully aware of the perils involved in applying the weak coupling methods at the scales on the 
order of $Q_s^2 \sim 1\ {\rm GeV}^2$. 
There are reasons to believe (see, e.g., \cite{KL}) that the gluon distributions at these scales may have a strong  
non--perturbative component, which would influence the results. The clarification of related problems, as well as 
the problem of subsequent evolution of the produced partons,  
await further work. 

We hope that the forthcoming experimental results and 
further progress in theory will 
eventually allow to uncover the dynamics of QCD at the high parton density, strong color field frontier.

\bigskip

We thank L. McLerran and R. Venugopalan for useful discussions of the results, and   
J.-P. Blaizot, K. Eskola, M.Gyulassy, K. Kajantie, Yu. Kovchegov, E. Levin, A.H.Mueller, 
H. Satz and X.-N. Wang for helpful conversations on the issues related to this 
work. The research of D.K. is supported by the U.S. Department of Energy under Contract No. DE-AC02-98CH10886.

\end{narrowtext}
\end{document}